\documentclass [twocolumn, aps, showpacs] {revtex4}
\usepackage{graphicx}
\usepackage{amsmath}
\usepackage{amssymb}
\usepackage{epsf}
\newlength{\upit}\upit=0.1truein

\newcommand{\ltappr}{{{\lower4pt\hbox{$<$} } \atop \widetilde{ \ \ \ }}}
\newlength{\bxwidth}\bxwidth=1.5 truein

\def\be{\begin{equation}}
\def\ee{\end{equation}}
\def\eqa{\begin{eqnarray}}
\def\eea{\end{eqnarray}}

\newlength{\figwidth}
\newlength{\shift}
\begin{document}
\title {Efficiency of Ground State Quantum Computer }

\author {Wenjin Mao}
\affiliation{ Department of Physics and Astronomy, Stony
Brook University, SUNY, Stony Brook, NY 11794, U.S.A.}
\date{\today}
\begin{abstract}
The energy gap is calculated for the ground state quantum computer circuit, which was recently proposed by Mizel et.al. When implementing a quantum algorithm by Hamiltonians containing only pairwise interaction, the inverse of energy gap $1/\Delta$ is proportional to $N^{4k}$, where $N$ is the number of bits involved in the problem, and $N^k$ is the number of control operations performed in a standard quantum paradigm. Besides suppressing decoherence due to the energy gap, in polynomial time ground state quantum computer can finish the quantum algorithms that are supposed to be implemented by standard quantum computer in polynomial time.
\end{abstract}
 \pacs{03.67.Lx}
\maketitle

Quantum computer is widely believed to outperform its classical counterpart for some classically difficult problems\cite{Shor, Grover, Farhi1, local}. Although many schemes have been suggested\cite{d1,d2,d3,d4,d5,d6,d7,d8}, the main obstacle is decoherence that makes very difficult the realization of even one qubit or one gate for quantum computing.  Ground state quantum computer (GSQC) is a new approach proposed by Mizel et.al.\cite{Mizel1, Mizel2, Mizel3}, which mimics time evolution of a system by space distribution of ground state wavefunction. 
The advantage of GSQC is that appreciable energy gap suppresses decoherence from environment as long as temperature is low enough.

To analyze the performance of GSQC, the key is  to evaluate  the scale of $\Delta$, the energy gap between the ground state $|\psi_0\rangle$ and the first excited state $|\psi_1\rangle$, which is related to the time cost. Mizel et.al. gave estimations on single qubit and two-qubit circuit in \cite{Mizel2}. However, they didn't  extend to arbitrary circuit with boost or projection Hamiltonians applied, while boost and projection Hamiltonians are necessary to make sure that $|\psi_0\rangle$ concentrates on the position corresponding to the final time in standard paradigm  so that measurement on GSQC can read out desired information with appreciable probability. In \cite{Mizel3} it was proposed to shorten length, for a single qubit, by inserting teleportation circuits, however, it will be shown that there the energy gap scale is incorrect. In the present paper, the scaling of energy gap is analyzed for general circuit that can be used for any known quantum algorithm. I will use boundary condition different  from \cite{Mizel3}, and show that there exist dangerous cases, in which the energy gap shrinks exponentially as the number of qubits increases. In order to prevent such exponential shrink, teleportation circuits are also inserted for multiple interacting qubits. Finally energy gap scaling for general quantum algorithm is presented.

 A standard computer is characterized by time dependent state as:
$
|\psi(t_i)\rangle=U_i|\psi(t_{i-1})\rangle,
$
where $t_i$ denotes instance of  the $i$-th step, and $U_i$ represents for unitary transformation. For GSQC, the time sequence is mimicked by  the space distribution of the ground state wavefunction $|\psi_0\rangle$. 
%The time cost is related with the inverse of energy gap $\Delta$.

As proposed by Mizel et.al.\cite{Mizel1}, a single qubit may be a column of  quantum dots with multiple rows, and each row contains a pair of quantum dots. State $|0\rangle$ or $|1\rangle$ is represented by finding electron in one of the two dots. It is important to notice that only one electron exists in a qubit. A GSQC is made up by circuit of multiple interacting qubits, whose ground state is determined by the summation of single qubit unitary transformation Hamiltonian $h(U_j)$, two-qubit interacting Hamiltonian $h(CNOT)$, boost Hamiltonian $h(B,\lambda)$ and projection Hamiltonian $h(|\gamma\rangle,\lambda)$. 

The single qubit unitary transformation Hamiltonian has the form:
\eqa
h^j(U_j)&=&\epsilon\left[ 
C^{\dagger}_{j-1}C_{j-1}+C^{\dagger}_{j}C_{j} \right.\nonumber\\
&& \ \ \ \ \ \ \ \ \ \left. -\left(C^{\dagger}_{j}U_jC_{j-1}+h.c.\right)\right],
\eea
where  $\epsilon$ defines the energy scale of all Hamiltonians, $C^{\dagger}_j=\left[c^{\dagger}_{j,0}\ c^{\dagger}_{j,1}\right]$, $c_{j,0}^{\dagger}$ is the electron creation operator on row $j$ at position $0$, and $U_j$ is two dimension matrix representing for unitary transformation from row $j-1$ to row $j$. The boost  Hamiltonian is:
\eqa
h^j(B,\lambda)&=&\epsilon\left[
C^{\dagger}_{j-1}C_{j-1}+\frac{1}{\lambda^2}C^{\dagger}_{j}C_{j}\right. \nonumber\\
&& \ \ \ \ \ \ \ \ \left.-\frac{1}{\lambda}\left(C^{\dagger}_{j}C_{j-1}+h.c.\right)\right],
\eea
which amplifies the wavefunction amplitude by large number $\lambda$ compared with previous row in $|\psi_0\rangle$. The projection Hamiltonian is
\eqa
h^j\left(|\gamma\rangle,\lambda\right)&=& \epsilon \left[
c^{\dagger}_{j-1,\gamma}c_{j-1,\gamma} +\frac{1}{\lambda^2}c^{\dagger}_{j,\gamma}c_{j,\gamma}  \right. \nonumber\\
&&\ \ \ \ \ \ \left. -\frac{1}{\lambda}\left(c^{\dagger}_{j,\gamma}c_{j-1,\gamma}+h.c.\right)\right], 
\eea
where $|\gamma\rangle$ represents for state to be projected to on row $j$ and to be amplified by $\lambda$. The interaction between qubit $\alpha$ and $\beta$ can be represented by $h(CNOT)$:
\eqa
h^j_{\alpha,\beta}(CNOT)&=&
\epsilon C^{\dagger}_{\alpha,j-1} C_{\alpha,j-1} C^{\dagger}_{\beta,j} C_{\beta,j}\nonumber\\
& &+h^j_{\alpha}(I)C^{\dagger}_{\beta,j-1} C_{\beta,j-1}
+ c^{\dagger}_{\alpha,j,0} c_{\alpha,j,0}h^j_{\beta}(I)\nonumber\\
& &
+c^{\dagger}_{\alpha,j,1} c_{\alpha,j,1}h^j_{\beta}(N). \label{hCNOT}
\eea
where for $c^{\dagger}_{a,b,c}$, its subscription $a$ represents for qubit $a$, $b$ for the number of row, $c$ for the state $|c\rangle$. 
%We will also use Hamiltonian $h\left(C(I)\right)$ that can be obtained by replacing $NOT$ to $I$ in Eq.(\ref{hCNOT}).
 With only $h^j(U_j)$ and $h^j_{\alpha,\beta}(CNOT)$, its ground state is\cite{Mizel2}:
\eqa
|\psi_0^j\rangle&=&\left[
1+c^{\dagger}_{\alpha,j,0}c_{\alpha,j-1,0}\left(1+C^{\dagger}_{\beta,j}C_{\beta,j-1}\right)\right.
\nonumber\\
&&\left. +c^{\dagger}_{\alpha,j,1}c_{\alpha,j-1,1}\left(1+C^{\dagger}_{\beta,j}NC_{\beta,j-1}\right)\right]\nonumber\\
&&\times \prod_{a\ne \alpha,\beta}\left(1+C^{\dagger}_{a,j}U_{a,j}C_{a,j-1}\right)|\psi^{j-1}\rangle.
\label{cnot}
\eea
All above mentioned  Hamiltonians are positive semidefinite, and are the same as those in \cite{Mizel1, Mizel2, Mizel3}. Only pairwise interaction is considered for Hamiltonians involving multiple qubits.

 The input states are determined by the boundary conditions applied upon the first rows of all qubits, which can be Hamiltonian $h^0=E(I+\sum_i a_i\sigma_i)$ with $\sigma_i$ being Pauli matrix and $\sum_ia_i^2=1$. For example, if $h^0=E(I+\sigma_z)$, then $|\psi_0\rangle$ on the first row is $|1\rangle$; if  $h^0=E(I-\sigma_x)$,  then  it is $ \left(|0\rangle+|1\rangle\right)/\sqrt{2}$. Unlike in \cite{Mizel3}, we set $E\gg \epsilon>0$, thus boundary Hamiltonians are not perturbation. If $E$ is large enough, the energy gap is independent of its magnitude. A typical value may be $E\approx 10 \epsilon$.

Although one can analytically obtain energy gap for single qubit with $n$ rows\cite{Mizel2}, 
it's difficult to calculate for a complicated circuit. However, we can still manage to find its scale. 
%Within this paper, the number of rows $n$ on each qubit is a small number for our purpose.

Let's first consider the simplest case:
 a $n$-row single qubit without projection or boost Hamiltonian. The Hamiltonian $\sum_{i=1}^n h(U_i)$ is in fact the kinetic energy just like a particle in a one-dimension box with length $n$, thus   $\Delta\propto \epsilon/n^2$. 
 The unique ground state $|\psi_0\rangle$ is uniform over all rows, and
 $|\psi_1\rangle$ is orthogonal to $|\psi_0\rangle$ on each row. Due to large value of $E$ in $h^0$, $|\psi_1\rangle$ on the first row is nearly zero, and it is an mono-increasing function that reaches maximum on the final row so as to keep wavefunction smooth. When a boost Hamiltonian $h(B,\lambda)$ is applied to  the  qubit's final row, $|\psi_1\rangle$ concentrates there, where the wavefunction amplitude is about $\lambda$ times those at other rows on average, so the wavefunction amplitude on other rows is $O(1/\sqrt{\lambda^2+n-1})$. When the qubit length $n\ll \lambda$, $|\psi_1\rangle$ is nearly a linear function of position except for at final row, and its amplitude at the first row is zero. The local kinetic energy is almost constant along the whole qubit, proportional to the square of difference of wavefunction between neighboring rows, hence energy gap scales as: 
\eqa
\Delta\propto \frac{\epsilon}{\lambda^2}.
\eea

%In \cite{Mizel2} $\Delta$ for a single qubit is found independent of $\lambda$ because they first found  the gap between doubly degenerated levels, and then obtained the real gap by perturbation theory, which might be right for small $\lambda$. However, for large $\lambda\gg n$ perturbation theory doesn't work, and this also happens to their energy gap calculation for multiple qubits\cite{Mizel2, Mizel3}.
%In our calculation we find that the energy gap between the second excited state and ground state is independent of $\lambda$.
%In this situation, it is claimed in \cite{Mizel2}  that the order of gap is not affected by $\lambda$, which is only correct in the absence of boundary condition.

If instead of boost Hamiltonian, a projection Hamiltonian $h(|\gamma\rangle,\lambda)$ is applied on the last row of a single qubit, then $|\psi_0\rangle$ on the final row is restricted to state $|\gamma\rangle$. 
Assuming besides amplitude and phase $|\psi_0\rangle$ on the second last  row is at state $|\xi\rangle$, which is normalized, and $ \langle \xi|\gamma\rangle$ is appreciable, then $|\psi_0\rangle$ will concentrate on the final row, and $|\psi_1\rangle$ should have little weight there, otherwise $\langle \psi_1|\psi_0\rangle\ne 0$. Thus $\Delta $ remains at about $\epsilon/n^2$, independent of $\lambda$. 

Numerical calculations on single qubit in various situations have confirmed all the above analysis. For example, a 6-row qubit with $h(B,\lambda)$ on final row and with boundary Hamiltonian, $h^0=10\epsilon(I-\sigma_z)$, on its first row, has energy gap $\Delta=0.0782,\ 0.0174,\ 1.94\times 10^{-3},\ 1.96\times 10^{-4},\ 1.96\times 10^{-5}\epsilon$ at $\lambda=1,\ \sqrt{10},\ 10,\ \sqrt{1000},\ 100$ respectively.

\begin{figure}
\begin{center}
\leavevmode
\hbox{\epsfxsize=6cm \epsffile{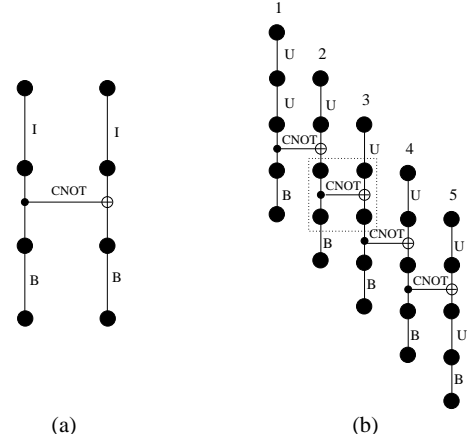}}
\end{center}
\caption{\small Part (a) shows two qubits interacting with each other by $h(CNOT)$. Part (b) shows a chain of qubits interacting with each other through $h(CNOT)$. The label $I$ stands for identity transformation, $CNOT$ for $h(CNOT)$, $U$ for arbitrary unitary transformation and $B$ for boost Hamiltonian $h(B,\lambda)$.}
\label{Two}
\end{figure}

 Next we consider two qubits interacting with each other through  $h(CNOT)$ as shown in Fig.(\ref{Two}a), with the left qubit as control qubit, the right one as target qubit, and both ending with $h(B,\lambda)$. By observing $|\psi_0\rangle$ in Eq.(\ref{cnot}), it's easy to see that it has the form 
\eqa
 &&\left( |\psi^{control}_{upstream}\rangle +|\psi^{control}_{downstream} \rangle \right) |\psi^{target}_{upstream}\rangle\nonumber\\
&&+ |\psi^{control}_{downstream}\rangle |\psi^{target}_{downstream}\rangle,
\label{wavefunction}
\eea
where the Hamiltonian $h(CNOT)$ divides both qubits into upstream and downstream parts with $h(B,\lambda)$ in downstream parts.
$h(B,\lambda)$ on the final row of the target qubit also raises the amplitude of $|\psi_0\rangle$ on the downstream rows of the control qubit, while $h(B,\lambda)$ on control qubit doesn't influence the amplitude on the target qubit at all because both parts of states entangled with the downstream state of control qubit. Thus on the control qubit the amplitude of upstream over its final row is $1/\lambda^2$, and it is $1/\lambda$ on target qubit because the state on final row of control qubit entangles with both upstream and downstream part of wavefunction on target qubit. Numerical calculation on $|\psi_0\rangle$ agrees exactly with this analysis.

The first excited state should be such a state that on the first row of one of the two interacting qubits, $|\psi_1\rangle$ is orthogonal to $|\psi_0\rangle$, and its amplitude is nearly zero due to large on-site potential from boundary Hamiltonian $h^0$, while on the first row of the other qubit $|\psi_1\rangle$ is at the same state as $|\psi_0\rangle$ with only amplitude modified. If one knows the overall amplitude of wavefunction in upstream part on the qubit whose first row  state is orthonormal to $|\psi_0\rangle$,
\eqa
\frac{1}{x}\approx \||\psi_1\rangle_{upstream}\|,
\label{X}
\eea
 then the scale of energy gap can be found by same reasoning for single qubit:
\eqa
\Delta\propto \epsilon/x^2.
\eea
If $|\psi_1\rangle$ maintains similar weight distribution to $|\psi_0\rangle$, we find that $|\psi_1\rangle$ should be orthogonal to $|\psi_0\rangle$ on the first row of the control qubit, in which $1/x\propto 1/\lambda^2$, while on target qubit $1/x\propto 1/\lambda$, thus the energy gap $\Delta\propto \epsilon/\lambda^4$. The larger value $1/x$ on target qubit doesn't determine the energy scale because there at the first row $|\psi_1\rangle$ has appreciable amplitude.

Numerical calculation agrees with above analysis for $|\psi_1\rangle$. We also found that the second excited state $|\psi_2\rangle$ has the same energy gap scaling as $|\psi_1\rangle$. $|\psi_2\rangle$ on the first row of target qubit is orthonormal to $|\psi_0\rangle$ with nearly zero amplitude, while its first row state on control qubit is the same as $|\psi_0\rangle$. $1/x$ on both target qubit and control qubit at $|\psi_2\rangle$ are of the same order $1/\lambda^2$, hence the energy gap is also  proportional to $\epsilon/\lambda^4$. The parameter $1/x$ on target qubit in $|\psi_2\rangle$ is different from those in $|\psi_0\rangle$ and $|\psi_1\rangle$ because $|\psi_2\rangle$ at the first row of target qubit has to be nearly zero, according to $h(CNOT)$, the suppression of amplitude of wavefunction requires larger difference over two neighboring rows to supply same amount of kinetic energy. In both $|\psi_1\rangle$ and $|\psi_2\rangle$, the value $1/x$ 
maintains the same order as ground state
on the qubit whose first row state is the same as $|\psi_0\rangle$.

If projection Hamiltonians are applied instead of  boost Hamiltonians, and at $|\psi_0\rangle$ $\langle \xi|\gamma\rangle$ is appreciable, where $|\xi \rangle$ is the state on the row before $h(B,\lambda)$, then $\Delta\propto \epsilon/\lambda^2$ due to the same reason as for single qubit. If there is one boost Hamiltonian and one projection Hamiltonian on the two interacting qubits, energy gap is still $\Delta\propto \epsilon/\lambda^4$. It is interesting to note that by projecting the target qubit into one state on its last row and boost this state, one can manipulate the state on the final row of the control qubit, and this can be observed in Eq.(\ref{wavefunction}). 

Numerical calculations confirm our analysis. For example, the energy gap for two 4-row qubits ended with $h(B,\lambda)$, as shown in Fig.(\ref{Two}a) with $h_1^0=10\epsilon(I-\sigma_x),\ h_2^0=10 \epsilon(I+\sigma_z)$, is $\Delta=0.0574,\ 2.43 \times 10^{-3},\ 3.05\times 10^{-5},\ 3.12\times 10^{-7},\ 3.13\times 10^{-9} \epsilon$ at $\lambda=1,\ \sqrt{10},\ 10,\ \sqrt{1000},\ 100$ respectively.

In the works by Mizel et.al.\cite{Mizel2}, when $h(B,\lambda)$ are applied to two interacting qubits, their lower bound energy gap agrees with result in the present paper, however, they didn't elaborate further for multiple interacting qubits. Fig.(7) in \cite{Mizel3} shows how to use teleportation\cite{book} to increase $\Delta$ for single qubit evolution, and there $\Delta$ is proportional to $\epsilon/\lambda^2$, which is incorrect because even two interacting qubits will give energy gap of $\epsilon/\lambda^4$ as shown in our analysis and numerical calculation for two qubit case, while the teleportation circuit involves more than two qubits interacting with each other. However, continuous single qubit evolution can always be combined into one unitary transformation, there is no need to evaluate this situation.

With multiple qubits interacting with each other, we need to evaluate on the top part of each qubit the parameter $1/x$ defined in Eq.(\ref{X}) assuming that only  on the first row of that qubit the excited state is orthonormal to $|\psi_0\rangle$ while states on all other qubits remains the same as corresponding ground state with only magnitude changed.
Then the minimum $1/x$ gives the energy gap scale as $\epsilon(1/x)_{min}^2$. Thus the energy gap depends on the detail of the circuit. 

In principle, when all qubits end with either projection or boost Hamiltonian containing same $\lambda$, at ground state and first excited state, the boost Hamiltonian or  projection Hamiltonian raises the amplitude of wavefunction on the final row by $\lambda$ compared without boost or projection Hamiltonian. When estimating $1/x$ of $|\psi_1\rangle$ for any qubit, say qubit $A$, the boost Hamiltonian, not the projection Hamiltonian, on qubit $A$ itself contributes $\lambda$ to $x$, and projection or boost Hamiltonians from its control operation mates contribute $\lambda$ to its $x$ value; contribution from those qubits not directly interacting with qubit $A$ can be analyzed according to ground state wavefunction distribution, Eq.(\ref{wavefunction}).

%\begin{figure}
%\begin{center}
%\leavevmode
%\hbox{\epsfxsize=2cm \epsffile{Chain.eps}}
%\end{center}
%\caption{\small A chain of qubits interact with each other through $h%%(CNOT)$. This is the circuit with the smallest energy gap.}
%\label{Chain}
%\end{figure}

For example, in Fig.(\ref{Two}b) qubit 1 $CNOT$ controls qubit 2, and in its downstream part, qubit 2 $CNOT$ controls qubit 3, so on to qubit 5. When all qubits end with $h(B,\lambda)$, we find that, for     $\lambda\gg 5$ ( 5 is the typical length of qubit) the parameter in Eq.(\ref{X}) on qubit 5 has contribution only from $h(B,\lambda)$'s on itself and qubit 4, those on qubit 3, 2 and 1 doesn't make contribution because on qubit 5 its upstream part can entangle with the final rows of other qubits except for qubit 4. We get $x\propto\lambda^2,\ \lambda^3,\ \lambda^4,\ \lambda^5, \lambda^5$ for qubit 5, 4, 3, 2 and 1 respectively. The energy gap is determined by value of $1/x$ on qubit 1 or 2, hence $\Delta\propto \epsilon/\lambda^{10}$. As number of qubits $N$ in such a chain increases, the energy gap shrinks as $\epsilon/\lambda^{2N}$.

The energy gap of a GSQC  circuit may be exponentially small depending on detail of circuits. The chain in Fig.(\ref{Two}b) is the most dangerous circuit giving the least energy gap, the other circuit that also suffers exponential shrink of energy gap $\Delta\propto\epsilon/\lambda^{2N}$ is a qubit interacting with $N$ other qubit as the control qubit. These two kinds of circuits are quite common in quantum algorithm, such as quantum Fourier transform.

\begin{figure}
\begin{center}
\leavevmode
\hbox{\epsfxsize=6cm \epsffile{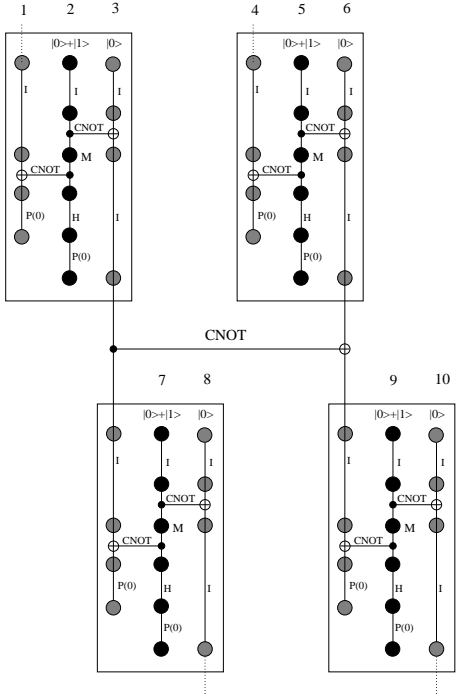}}
\end{center}
\caption{\small Modified circuit of the dotted box in Fig.(\ref{Two}b), in which four teleportation boxes are inserted before and after $H(CNOT)$. The circuit inside each teleportation box is similar to the teleportation circuit in \cite{Mizel3}. Label $I$ represents for identity transformation Hamiltonian $h(I)$, $H$ for Hadamard transformation Hamiltonian $h(H)$ and $P(0)$ for projection Hamiltonian $h(|0\rangle,\lambda)$. The dotted lines represent for part of circuit not showing.}
\label{detail}
\end{figure}
In order to break such exponential shrink of $\Delta$ like  in Fig.(\ref{Two}b), the circuit is modified by inserting teleportation boxes for all qubits between two control Hamiltonians, as in Fig.(\ref{detail}), which shows how the circuit in the dotted box of Fig.(\ref{Two}b) is modified. A teleportation box teleports quantum state of a qubit to the other one, and it can be implemented in both standard paradigm\cite{book} and GSQC\cite{Mizel3}. The whole chain in Fig.(\ref{Two}b) should be modified similarly. In Fig.(\ref{detail}), the four boxes contain teleportation circuits similar to that in \cite{Mizel3}. On the middle qubit of teleportation boxes in Fig.(\ref{detail}), the row labeled by $M$ opens a gate for its left side qubit's upstream part to entangle with downstream rows at its right side qubit, thus the exponential shrink of $1/x$ along the chain is prevented. The minimum $1/x$ happens on qubit 3 or qubit 6 in Fig.(\ref{detail}). For example, on qubit 3, the parameter $1/x$ has contributions only from projection Hamiltonians $h(|0\rangle,\lambda)$ on qubit 1, 2, 6 and  7, hence $1/x\propto 1/\lambda^4$. While projection or boost Hamiltonian on qubit 8 or 10 doesn't contribute to $1/x$ on qubit 3 because through rows marked by $M$ on the middle qubit in teleportation boxes, their final row states entangle with the first two-row state of qubit 3, and they surely also entangle with the last row state on qubit 3. Qubit 9 doesn't contribute to $1/x$ on qubit 6, thus has no effect on qubit 3 because the  boost or projection Hamiltonian on control qubit doesn't affect the parameter on target qubit at ground state, while the lowest excited state remains similar with ground state on qubit 9 if only on qubit 3 the first row state is orthonormal to $|\psi_0\rangle$. Just like qubit 8 and 10, other qubits further away that do not show up in Fig.(\ref{detail}) do not affect $1/x$ on  qubit 3. By inserting teleportation boxes, we find that when the amplifying factor of all boost and projection Hamiltonians is $\lambda$, the energy gap for arbitrary GSQC circuit is always 
\eqa
\Delta\propto\epsilon/\lambda^8.
\eea
Hence there is no chain action on $1/x$ like in Fig(\ref{Two}b) that leads to exponential shrink of energy gap. In fact, the upper right and the lower left teleportation boxes are not needed concerning on circuit like Fig.(\ref{Two}b), however, I keep them for more general consideration.

For arbitrary circuit, it can always be modified similar to Fig.(\ref{detail}) and the energy gap will be kept at the scale of $\epsilon/\lambda^8$. The price of this modification is that the total number of qubits increases, thus from measurement concern, it requires larger value of $\lambda$ to make sure appreciable probability to find all electrons on the final row's of all qubits. However, this price is worthy because it may change energy gap from exponentially small to polynomially small, while the value of $\lambda$ is only polynomially increased. This can be demonstrated in the following example.

%on the right qubit in some teleport boxes with $1/x\approx 1/\lambda^4$, where the other two qubits in the same teleport box contribute $\lambda^2$, and qubits in the next two teleport boxes on the right side in the chain contribute $\lambda^2$. Those qubits further away from it have no influence upon its $1/x$ value. If there are $L$ qubits forming a chain like in Fig.(\ref{Two}b), and all ended with $h(B,\lambda)$, then the energy gap is $\Delta\propto \epsilon/\lambda^{2L}$, while the modified circuit by inserting teleport box will have energy gap $\Delta\propto \epsilon/\lambda^{8}$. Other circuit can be analyzed similarly, and we find that $\epsilon/\lambda^8$ is the minimum energy gap.

Now we can  check the energy gap when a quantum algorithm is implemented by GSQC. The most powerful quantum algorithm up to now is quantum Fourier transform, which is used in factorization, period finding, etc. The detail of quantum Fourier transform, more precisely, the inverse quantum Fourier transform, can be found in many literatures, such as \cite{book}. It takes $O(N^2)$ control operations $h(CR^{\dagger}_k)$ to carry out inverse quantum Fourier transform by standard time dependent approach, where $N$ is the number of qubits involved in the problem, and $R^{\dagger}_k$ is  the unitary transformation shifting phase of state $|1\rangle$ by $-2\pi i/2^k$\cite{book}. $h(CR^{\dagger}_k)$ can be obtained by replacing $NOT$ operator by $R^{\dagger}_k$ in Eq.(\ref{cnot}). One can simply replace the time evolution by circuit array and calculate the energy gap of the circuit. By inserting teleportation boxes between control operations, when all qubits end either by $h(|0\rangle,\lambda)$ or $h(B,\lambda)$, the energy gap is $
\Delta\propto {\epsilon}/{\lambda^{8}}$.
Because after experiencing a control operation each qubit is teleported to a new qubit, the total number of qubits is of the order $N^2$ instead of $N$. With each qubit ended with boost or projection Hamiltonian, at $|\psi_0\rangle$ the probability of finding electron on the final row on any qubit is larger than $(1-C/\lambda^2)$ with $C\le 8$ being the number of rows on the qubit. In order to find all electrons on final rows with appreciable probability $P\ge(1-C/\lambda^2)^{FN^2}$, it requires $\lambda\approx \sqrt{D} N$, here we assume there are totally $FN^2$ qubits involved, and $D$ is a constant that could be tuned. Thus the probability is $P\approx e^{-FC/D}$ as $N$ being large number, the energy gap is $\Delta\propto \epsilon/(D^4N^{8})$, noting that standard paradigm needs only $O(N^2)$ steps. 

We conclude that the energy gap of ground state quantum computer is determined by the number of control operation, and any quantum algorithm implemented by standard paradigm can be implemented by GSQC, whose energy gap is  $\Delta\propto 1/N^{4k} $, where $N$ is the number of bits in the problem, and $N^k$ is the number of control operations needed in standard quantum paradigm. There are various approaches to make GSQC stay at it's ground state, and the time cost is determined by different approaches. Estimation of time can be obtained by adiabatic approach, in which we increase $\lambda$ slowly from 1 to $O(N^{k/2})$ in all boost and projection Hamiltonians on final row of all qubits, GSQC will stay at ground state and time cost is of the order $N^{8k} $, determined by inverse of square of minimum energy gap.
By local adiabatic approach, the time cost is even shorter\cite{local}.

%Quantum algorithms implemented by standard paradigm in polynomial time may be implemented by ground state quantum computer in polynomial time, and the energy gap may suppress environment decoherence, hence GSQC might be more feasible than other forms of quantum computer. On the other hand, while it doesn't make sense to implement Grover's algorithm by GSQC, there are some algorithms that standard paradigm cannot realize but  GSQC might  capable, and we will report those algorithms in following work.

I would like to thank A. Mizel for helpful discussion and for pointing out the mistake in early version of the manuscript. This
work is supported in part by the NSF under grant \# 0121428 and
by ARDA and DOD under the DURINT grant \# F49620-01-1-0439.


\begin{thebibliography}{10}
\small
\bibitem{Shor}P. Shor, in {\textit{Proceedings of the 35th Annual Symposium on the Foundations of Computer Science, Los Alamitos, California, 1994}}, edited by Goldwasser (IEEE Computer Society Press, New York, 1994), p. 124.
\bibitem{Grover}L.K. Grover, Phys. Rev. Lett. {\bf 79}, 325(1997).
\bibitem{Farhi1}E. Farhi, J. Goldstone, S. Gutmann, J. Lapan, A. Lundgren, D. Preda, Science,  {\bf{292}}, 472(2001).
\bibitem{local} J. Roland and N.J. Cerf, Phys. Rev. A {\bf 65}, 042308(2002).
\bibitem{d1} Q. A. Turchette {\it et al.} {\em Phys. Rev. Lett.} {\bf 75}, 4710 (1995).
\bibitem{d2} I. L. Chuang, N. Gershenfeld, M. Kubinec, {\em Phys. Rev. Lett.} {\bf 80}, 3408 (1998).
\bibitem{d3} D. Loss and D. P. DiVincenzo, {\em Phys. Rev. A} {\bf 57}, 120 (1998).
\bibitem{d4} G. Burkhard, D. Loss, D. P. DiVincenzo, {\em Phys. Rev. B} {\bf 59}, 2070 (1999).
\bibitem{d5} Y. Makhlin, G. Schon, A. Shnirman, {\em Nature} {\bf 398},
305 (1999).
\bibitem{d6} D. V. Averin, {\em Solid State Commun.} {\bf 105}, 659 (1998).
\bibitem{d7} B. Kane, {\em Nature} {\bf 393}, 133 (1998).
\bibitem{d8} L. B. Ioffe, V. B. Geshkenbein, M. V. Feigel'man, A. L.
Fauchere, G. Blatter, {\em Nature} {\bf 398}, 679 (1999).
\bibitem{Mizel1} A. Mizel, M.W. Mitchell and M.L. Cohen, Phys. Rev. A, {\bf 63}, 040302(2001).
\bibitem{Mizel2} A. Mizel, M.W. Mitchell and M.L. Cohen, Phys. Rev. A, {\bf 65}, 022315(2002).
\bibitem{Mizel3} A. Mizel, Phys. Rev. A, {\bf 70}, 012304(2004).
%\bibitem{teleport}M.A. Nielsen and I.L. Chuang, Phys. Rev. Lett. {\bf 79}, 321(1997).
\bibitem{book}M.A. Nielsen and I.L. Chuang, {\textit{ Quantum Computation and Quantum Information}}, Cambridge University Press, 2000.
\end{thebibliography}
\end{document}